\documentclass[%
 reprint, 
 superscriptaddress,
 amsmath,amssymb,
 aps,
 pre
]{revtex4-2}

\usepackage{graphicx}
\usepackage{dcolumn}
\usepackage{bm}
\usepackage{physics}
\usepackage{amsmath}
\usepackage{hyperref}
\hypersetup{colorlinks,allcolors=black}
\begin{document}


\title{Hamiltonian-Driven Architectures for \\ 
        Non-Markovian Quantum Reservoir Computing}

\author{Daiki Sasaki}
 \email{dsasaki@uec.ac.jp}
\author{Ryosuke Koga}
\author{Taihei Kuroiwa}
\author{Yuya Ito}
 \affiliation{Engineering Department of the University of Electro-Communications, 
	182-8585, Chofu, Tokyo, Japan}
\author{Chih-Chieh Chen}
 \affiliation{Grid Inc., 107-0061, Tokyo, Japan}
\author{Tomah Sogabe}

 \affiliation{Engineering Department of the University of Electro-Communications, 
	182-8585, Chofu, Tokyo, Japan}
 \affiliation{Grid Inc., 107-0061, Tokyo, Japan}
 \affiliation{i-PERC, The University of Electro-Communications, 
	182-8585, Chofu, Tokyo, Japan}
 \email{sogabe@uec.ac.jp}


\begin{abstract} 
We propose a Hamiltonian-level framework for non-Markovian quantum reservoir computing directly tailored for analog hardware implementations. By dividing the reservoir into a system block and an environment block and evolving their joint state under a unified Hamiltonian, our architecture naturally embeds memory backflow by harnessing entanglement-induced information backflow with tunable coupling strengths. Numerical benchmarks on short-term memory tasks demonstrate that operating in non-Markovian regimes yields significantly slower memory decay compared to the Markovian limit.  Further analyzing the echo-state property (ESP), showing that the non-Markovian quantum reservoir evolves from two different initial states, they do not converge to the same trajectory even after a long time, strongly suggesting that the ESP is effectively violated. Our work provides the first demonstration in quantum reservoir computing that strong non-Markovianity can fundamentally violate the ESP, such that conventional linear-regression readouts fail to deliver stable training and inference. Finally, we experimentally showed that, with an appropriate time-evolution step size, the non-Markovian reservoir exhibits superior performance on higher-order nonlinear autoregressive moving-average(NARMA) tasks.
\end{abstract}

\maketitle

\section{Introduction}

Reservoir computing (RC) is a neural-inspired paradigm in which a fixed, high-dimensional dynamical system, i.e., the reservoir, projects time-series inputs into a rich feature space while only a simple linear readout is trained \cite{Jaeger2001EchoState, Maass2002RealTime}. The key advantage is that only the readout weights are adjusted, leaving the reservoir’s internal dynamics unaltered, which enables rapid training and harnesses complex dynamics for efficient timeseries prediction and energy‐efficient edge‐device computing \cite{yan_emerging_2024}.

Quantum reservoir computing (QRC) extends this concept by using a quantum dynamical system as the reservoir, leveraging the richer dynamics and potentially exponential state space of quantum systems \cite{Fuji-Naka, Palacios_2024}. This allows QRC to inherently process quantum inputs and explore a vastly larger Hilbert space compared to classical reservoirs \cite{Palacios_2024}. Fujii and Nakajima introduced QRC as a way to “exploit natural quantum dynamics of ensemble systems for machine learning,” demonstrating that small quantum systems (on the order of 5–7 qubits) can emulate nonlinear dynamical systems and achieve performance comparable to classical recurrent neural networks of hundreds of nodes \cite{Fuji-Naka}. Such results opened a new avenue for analog quantum computing, where quantum physics powers tasks like time-series prediction and pattern recognition without requiring error-corrected quantum computers \cite{kornjača2024largescalequantumreservoirlearning, dudas_quantum_2023}.

\begin{figure*}[btp]
\centering
\includegraphics[scale=0.63]{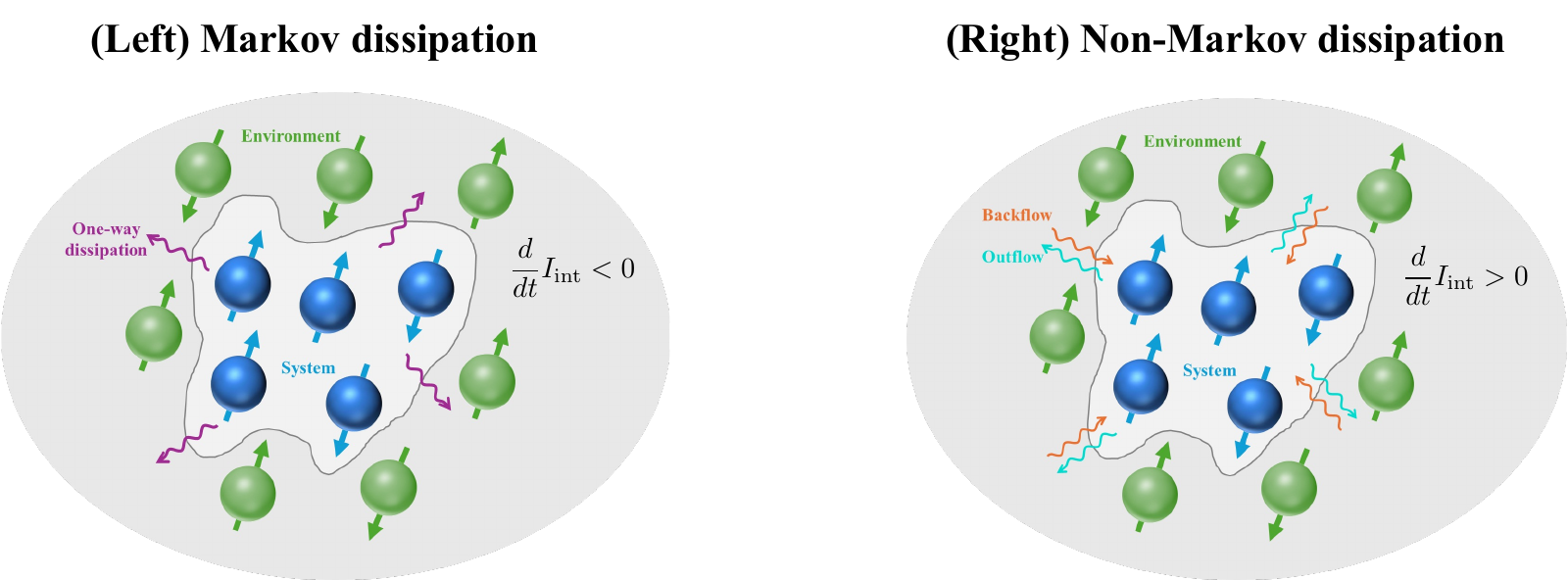}
\caption{Schematic comparison of information flow in Markovian vs. non-Markovian quantum reservoir dynamics. Left (Markovian): System qubits (blue) undergo one-way dissipation into the unobserved environment (green), indicated by purple arrows. No information returns to the system, so the total internal information $I_\mathrm{int}$ decays monotonically ($\frac{d}{dt}I_\mathrm{int}<0$). Right (Non-Markovian): Coherent backflow (orange arrows) from the environment partially restores system information, superimposed on the usual outflow (blue arrows). This bidirectional exchange causes intervals of increasing internal information ($\frac{d}{dt}I_\mathrm{int}>0$), enabling longer memory retention in the reservoir. Here,  denotes the information inside the open system, quantified by the trace distance between two reservoir states, }
\label{fig:spin}
\end{figure*}

Earlier works regarding QRC models primarily focused on straightforward open-system dynamics . The reservoir was often modeled as a set of qubits or quantum oscillators evolving under a fixed Hamiltonian (unitary dynamics), with some weak decoherence or regular resets to ensure a fading memory.  This mirrors the echo state property in classical reservoirs, where the influence of old inputs eventually decays. Indeed, achieving a balance between memory and forgetting is essential because the reservoir must remember input history enough to be useful, but not so long that it never forgets past data. In classical reservoir computing, rich dynamics (like chaos or nonlinearity) and short-term memory are crucial resources. Similarly, in QRC, quantum properties such as coherence, entanglement, and accessible memory are key to performance. Traditionally, researchers ensured this by either periodically partial re-initializing the quantum state at each time-step, or by assuming Markovian (memoryless) dissipation that continuously damps the system \cite{Kobayashi2024FDQRC, murauer2025feedbackconnectionsquantumreservoir, Mujal_2023, yasuda2023quantumreservoircomputingrepeated, Molteni_2023, Sannia2024dissipationas}.
It has been shown that these approaches effectively treat the reservoir’s evolution as Markovian or unitary – i.e. the future state depends only on the current state (or input) and not on any unmodeled history. Most QRC works mentioned before either assumed ideal unitary evolution or included only Markovian noise models (e.g. simple Lindblad damping) to introduce fading memory \cite{Sannia2024dissipationas}. Under these assumptions, the reservoir’s internal memory of past inputs decays exponentially with time. Therefore, standard QRC architecture inherently loses long-term information, limiting their performance on tasks requiring long-range temporal correlations. 

Non-Markovian dynamics – where the reservoir’s environment has a memory of past states – were largely overlooked. In an open quantum system, non-Markovianity refers to the presence of memory in the environment’s influence on the system. Formally, a quantum process is non-Markovian if the future evolution of the system depends on its earlier history, contrary to Markovian processes where the environment’s effect at any instant is independent of the past \cite{Breuer_2016, Rivas_2012, Vega_NM_2017} . Over the past decade, the theory of quantum non-Markovian dynamics has matured, with measures to quantify memory effects \cite{Liu_2011} and experimental techniques to control the degree of environmental memory \cite{BLP2009}. 
The consensus from this body of work is that information can flow back from the environment to the system in a non-Markovian regime as shown in Fig.~\ref{fig:spin}, effectively extending the system’s memory. In the context of QRC, this raises an enticing possibility: by engineering the reservoir’s dynamics to be non-Markovian, one might extend the memory depth of the reservoir and improve its capability on long-horizon temporal tasks.

Moreover, for true analog quantum hardware such as coupled spin networks, superconducting circuits, or cold-atom lattices, a direct Hamiltonian design offers a more natural and resource efficient route. In particular, for non-Markovian dynamics, characterized by information backflow from environment. Existing QRC schemes have not fully leveraged first-principles Hamiltonian engineering to harness these non-Markovian effects in a controlled manner.
In this work, we introduce Hamiltonian-driven architectures for non-Markovian QRC that explicitly specify the system Hamiltonian, environment Hamiltonian, and interaction Hamiltonian. By tuning the strength and spectral properties of their couplings, one can smoothly interpolate between Markovian and non-Markovian regimes and thereby design reservoirs with tailored memory kernels. We validate our approach on standard short-term memory benchmarks, quantifying performance via the coefficient of determination $R^2$ and showing markedly slower decay in non-Markovian settings. We further examine the echo-state property, revealing that strong environment coupling can break the usual reservoir contraction condition and lead to persistent oscillations in state trajectories. 
Finally, we assessed the performance dynamics of the proposed model on NARMA tasks that incorporate higher-order delays. By adjusting the time-evolution step size, we experimentally demonstrated that the non-Markovian reservoir exhibits superior performance on higher-order NARMA tasks.

\begin{figure*}[t]
\centering
\includegraphics[scale=0.68]{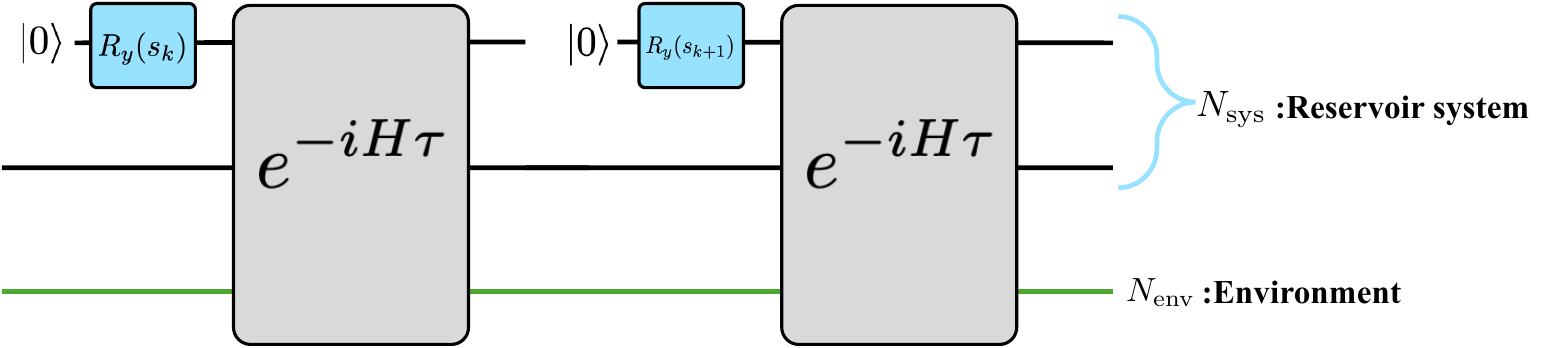}
\caption{Quantum‐circuit representation of the Hamiltonian-driven QRC time evolution. Blue boxes labeled $R_y(s_k)$ denote 
y-axis rotation gates that encode the input signal $s_k$ into each reservoir qubit. Gray blocks $e^{-iH\tau}$  implement joint unitary evolution of all qubits under the Hamiltonian $H$ for a time step width $\tau$. The top $N_{sys}$  wires (black) carry the reservoir system of 
$N_{sys}$  qubits, and the bottom 
$N_{env}$ wire (green) carries the unmeasured environment of 
$N_{env}$  qubits. Here, $N_{sys}$ and 
 $N_{env}$  abbreviate “number of system qubits” and “number of environment qubits,” respectively.}
\label{fig:qc}
\end{figure*}

The remainder of this paper is organized as follows. Section 2 formulates the Hamiltonian-driven QRC model and discusses its theoretical underpinnings. 
Section 3 demonstrates that, in reservoirs with strong non-Markovianity, the correlation with past inputs remains low over extended periods in the short-term memory task, exhibiting fragmented long-term memory.
Section 4 analyzes stability and the echo-state property. 
Section 5 illustrates the performance progression of the proposed model on higher-order NARMA tasks.
Finally, we present our conclusions and discuss future directions for non-Markovian quantum reservoirs.

\vspace{12pt}

\section{Model}
\subsection{Open Quantum Dynamics and Reservoir States}
We propose a model of quantum reservoir computing that extends the scheme introduced in \cite{Fuji-Naka}.  In our model, among the $N$ available qubits, $N_{\mathrm{sys}}$ qubits serve as the reservoir system, while $N_\mathrm{env}$ qubits form an unobservable environment. In other words, we consider a many-body spin system exposed to a spin bath \cite{Yang2009_spinbath, Krovi_2007}. This construction allows the natural Hamiltonian-driven time evolution to cover dissipation ranging from Markovian to non-Markovian.  Moreover, in the special case $N_\mathrm{env}=0$, the model reduces to the Fujii–Nakajima scheme \cite{Fuji-Naka}.

Further, information backflow arising from non-Markovianity causes the environment to act as a “reservoir of the reservoir,” which is expected to realize long-term memory.  Such backflow is absent in purely Markovian dynamics, where information dissipates unidirectionally (see Fig.~\ref{fig:spin}).

We consider the full quantum system of all $N$ qubits evolving under the Hamiltonian $H$:
\begin{equation}
H = (H_{\mathrm{sys}} \otimes I_{\mathrm{env}}) + (I_\mathrm{sys} \otimes H_{\mathrm{env}}) + H_\mathrm{int}
\end{equation}
Here, $H_{\mathrm{sys}}$ is the system Hamiltonian, $H_{\mathrm{env}}$ is the environment Hamiltonian, given by
\begin{align}
H_{\mathrm{sys}} &= \sum_{i<j}^{N_\mathrm{sys}} J_{ij}^{\mathrm{(sys)}} \sigma_i^x \sigma_j^x + h^{\mathrm{(sys)}} \sum_{i=1}^{N_\mathrm{sys}} \sigma_i^z, \\
H_{\mathrm{env}} &= \sum_{k<l}^{N_\mathrm{env}} J_{kl}^{\mathrm{(env)}} \sigma_k^x \sigma_l^x + h^{\mathrm{(env)}} \sum_{k=1}^{N_\mathrm{env}} \sigma_k^z.
\end{align}
The system–environment interaction Hamiltonian is
\begin{equation}
H_\mathrm{int} = \sum_{i=1}^{N_\mathrm{sys}} \sum_{k=1}^{N_\mathrm{env}} g_{ik} \sigma_i^z \sigma_k^z.
\end{equation}
The coupling coefficients are sampled as
\begin{align}
J_{ij}^{\mathrm{(sys)}} &\sim \mathrm{Uniform}(-J_0, J_0), \\
J_{kl}^{\mathrm{(env)}} &\sim \mathrm{Uniform}(-\alpha J_0, \alpha J_0), \\
g_{ik} &\sim \mathrm{Uniform}(-\beta J_0, \beta J_0).
\end{align}
Using $J_0$ as a reference scale, the parameters $\alpha$ and $\beta$ control the rate of information dissipation within the environment relative to the system, and the strength of the system–environment coupling, respectively.  This allows one to traverse between the Markovian dissipation regime $(\alpha\gg1,\ \beta\ll1)$ and the non-Markovian dissipation regime $(\alpha\ll1,\ \beta\gg1)$.

\begin{figure*}[t]
\centering
\includegraphics[scale=0.68]{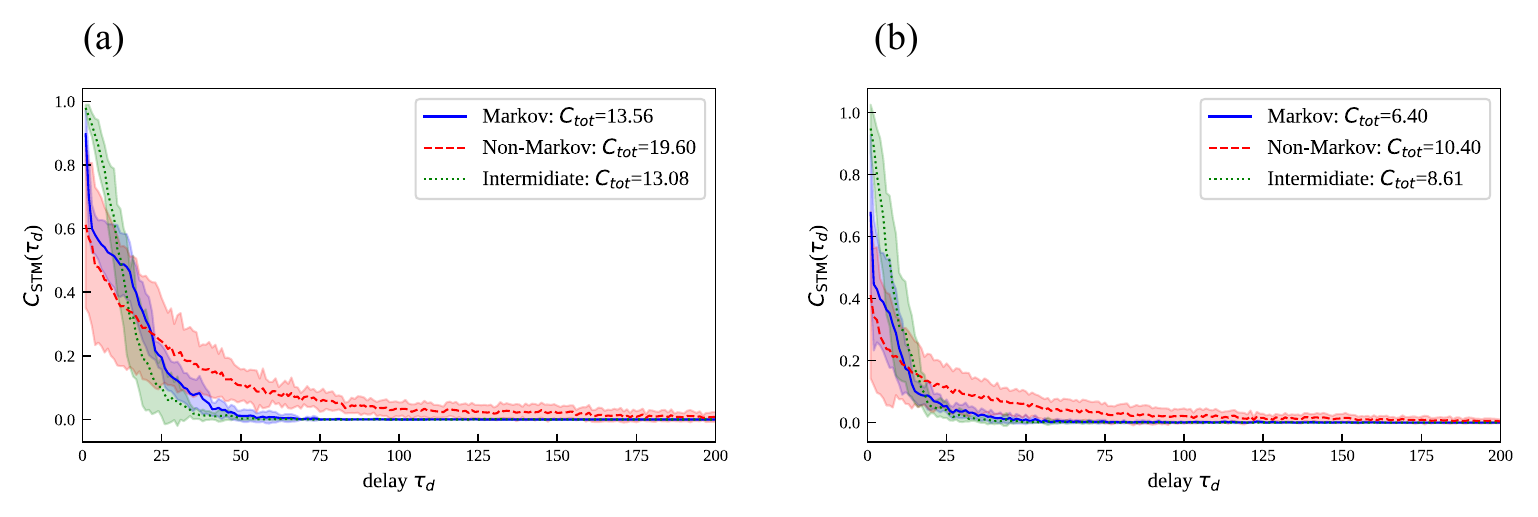}
\caption{Evolution of $C_{\mathrm{STM}}(\tau_d)$ in the short-term memory task. (a) $N=7, N_\mathrm{sys}=4, N_\mathrm{env}=3$. (b) $N=7, N_\mathrm{sys}=3, N_\mathrm{env}=4$. The parameter sets $(\alpha, \beta)=(10.0, 0.01), (0.01, 10.0), (1.0, 1.0)$ correspond to the ``Markov regime'', ``Non-Markov regime'', and ``Intermediate regime'', respectively.}
\label{fig:mc_7_4_3}
\end{figure*}

The time evolution of the full system state in small time step width $\tau$ is described by
\begin{equation}
\rho_{k+1} = e^{-i H \tau}\Bigl[\rho^\mathrm{(in)}_{k+1} \otimes \mathrm{Tr}_{\mathrm{in}}[\rho_k]\Bigr]e^{i H \tau}.
\end{equation}
Here, $\rho^\mathrm{(in)}_k := \ket{\psi_{s_k}}\bra{\psi_{s_k}}$ with $\ket{\psi_{s_k}} := \sqrt{1-s_k}\ket{0} + \sqrt{s_k}\ket{1}$.  The parameter $s_k\in[0,1]$ is the signal input to the reservoir at step $k$.  Since the environment remains inaccessible during evolution, we treat
\begin{equation}
\rho^\mathrm{(sys)}_k = \mathrm{Tr}_\mathrm{env}[\rho_k]
\end{equation}
as the effective system state.  This dynamics can be represented by the quantum circuit shown in Fig.~\ref{fig:qc}.
The reservoir readout vector at each step $k$ is defined by the expectation values of operators $O_i$:
\begin{equation}
x_{ki} = \mathrm{Tr}[\rho^\mathrm{(sys)}_k O_i]
\label{eq:qrc-res}
\end{equation}
In this paper, unless otherwise noted, we take $\{O_i\}_i = \{Z_i\}_{i=1}^{N\mathrm{sys}}$, or additionally the inter-site correlations $\{O_{ij}\}_{i<j} = \{Z_i Z_j\}_{i<j}$, where $Z_i$ and $Z_j$ denote the Pauli $Z$ measurement for the qubit $i$ and qubit $j$.

\subsection{Learning Readout Function}
We consider a supervised learning task in which a training set of input–output pairs $\{s_k, y_k\}_{k=1}^{L\mathrm{train}}$ is given, and the goal is to predict $y_k$ from $s_k$.  We regard the reservoir readout $x_{ki} = \Tr\bigl[\rho_k^{(\mathrm{sys})}\,O_i\bigr]$
as a feature vector that adequately captures temporal correlations, and seek weights $\{w_i\}$ that minimize the mean squared error
\begin{equation}
\frac{1}{L_\mathrm{train}}\sum_{k=1}^{L_\mathrm{train}}\bigl(y_k - \hat y_k\bigr)^2
\end{equation}
under the linear model
\begin{equation}
\hat y_k \;:=\;\sum_i w_i\,x_{ki}.
\end{equation}
These weights can be computed efficiently via the Moore–Penrose pseudoinverse. Although we have described a simple linear regression, other readout schemes—such as Ridge regression, multilayer perceptrons, or attention-based models—may likewise be employed.

\begin{figure*}[t]
\centering
\includegraphics[scale=0.80]{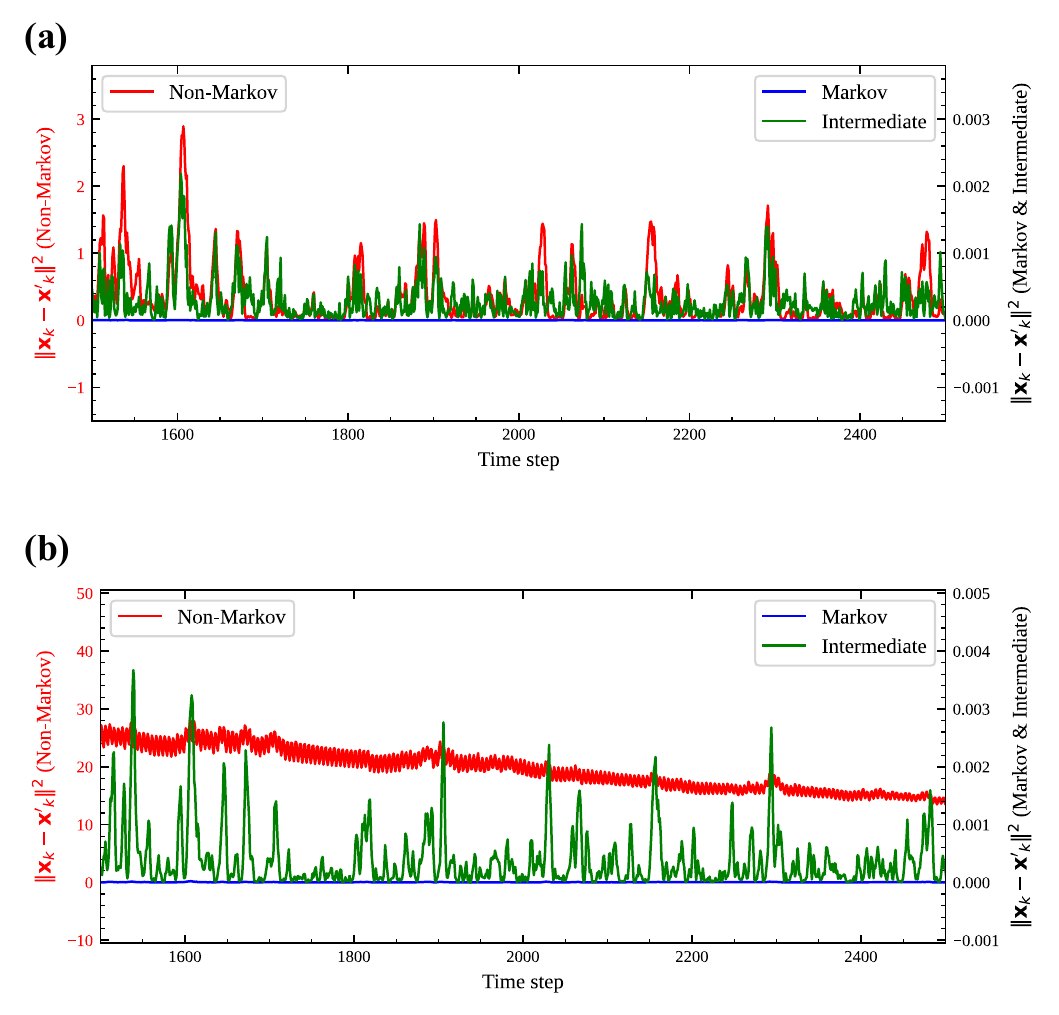}
\caption{Plot of the evolution of the squared norm of the difference between two reservoir‐state vectors, starting from two different quantum states and after sufficient updates. In panels (a) and (b), different random seeds were used during reservoir design (i.e., different realizations of the Hamiltonian interaction coefficients). These figures show the evolution of squared‐norm difference from step 1500 to 2500.}
\label{fig:esp_7_4_3}
\end{figure*}

\section{Residual Fragmented Memory Due to Non-Markovianity}

\subsection{Short Term Memory (STM) Task}
We consider the task of predicting past input signals $\{y_k(\tau_d)=s_{k-\tau_d}\}_k$ from the input sequence $\{s_k\}_k$.  In this delayed‐reproduction task, we denote the reservoir state and the predicted output by
$x_{ki}(\tau_d),\quad \hat{y}_k(\tau_d)$,
respectively.  Using the trained weights $w_i$, we compute the coefficient of determination ($R^2$-score) between the prediction $\hat{y}_k(\tau_d)$ and the true output $y_k(\tau_d)$ as
\begin{equation}
C_{\mathrm{STM}}(\tau_d) = \frac{\mathrm{Cov}^2(\bm{y}(\tau_d), \bm{\hat{y}}(\tau_d))}{\mathrm{Var}(\bm{y}(\tau_d))\mathrm{Var}(\bm{\hat{y}}(\tau_d))}.
\end{equation}
We evaluate $C_{\mathrm{STM}}(\tau_d)$ using validation data that were not used during training.

\subsection{Results}
Figure~\ref{fig:mc_7_4_3} shows the evolution of the coefficient of determination for short term memory $C_{\mathrm{STM}}(\tau_d)$ as a function of delay 
$\tau_d$ under three dynamical regimes, performing the STM task using the reservoir state updated by Eq.~(\ref{eq:qrc-res}).

 In Fig.~\ref{fig:mc_7_4_3}(a), we set $N=7$ with $N_\mathrm{sys}=4$ and $N_\mathrm{env}=3$, and in Fig.~\ref{fig:mc_7_4_3}(b), we set $N=7$ with $N_\mathrm{sys}=3$ and $N_\mathrm{env}=4$. The input signal $s_k$ is sampled independently at each time step from $\mathrm{Uniform}(0,1)$. We used 1000 steps for the washout period, 3000 steps for the training dataset, and 1000 steps for the validation dataset.
The parameter sets $(\alpha, \beta) = (10.0, 0.01), (0.01, 10.0), (1.0, 1.0)$ were employed to adjust the strengths of the intra-environment and system–environment interactions, corresponding respectively to the “Markov regime” (blue line), “Non-Markov regime” (red line), and “Intermediate regime” (green line). The intra-system coupling strength was fixed at $J_0=1$, and the magnetic fields were set to $h^\mathrm{(sys)} = J_0/2$ and $h^\mathrm{(env)} = \alpha J_0$. The time-evolution step width was $\tau = 0.5$, and the number of virtual nodes for time multiplexing was $V=50$ \cite{Fuji-Naka}.
As reservoir observables, we used the single-qubit expectation values $Z_i$. In this case, the dimension of the reservoir state vector at each time step is $N_x = V N_\mathrm{sys}$. Since we include a bias term during linear regression, the effective dimension of the reservoir state vector becomes $V N_\mathrm{sys} + 1$. Linear regression was used for the readout. The lines and shaded regions represent the mean and standard deviation, respectively, computed over ten different random seeds.
Under strong non-Markovian coupling, 
$C_{\mathrm{STM}}(\tau_d)$ decays markedly more slowly than in the Markovian or intermediate cases, indicating that fragments of past inputs persist for longer delays. Comparing Fig.~\ref{fig:mc_7_4_3}(a) and Fig.~\ref{fig:mc_7_4_3}(b), increasing environment size (from 3 to 4 qubits) further enhances long‐delay memory under non-Markovian dynamics.

\section{Stationary Echo State Property Under Non-Markovian Dissipation}

\subsection{Echo State Property in QRC}
We assume the reservoir state is updated according to
\begin{equation}
\bm{x}_\mathrm{next} = \bm{F}(\bm{x}, \bm{s}).
\end{equation}
For any input sequence $\bm{s}$ and any two distinct states $\bm{x}, \bm{x}'$,  consider
\begin{equation}
\|\bm{F}(\bm{x}, \bm{s})- \bm{F}(\bm{x}', \bm{s})\| \le r\|\bm{x} - \bm{x}'\|.
\end{equation}
When $0 \le r < 1$ holds, the echo state property (ESP) is said to be satisfied.  In other words, starting from two different reservoir states $\bm{x}$ and $\bm{x}'$, repeated updates will asymptotically converge to the same state—this is the stability condition of the reservoir.

In quantum reservoir computing, if one uses the density matrix itself (or equivalently the full set of Pauli‐string expectation values) as the reservoir state, ESP means that for the same input sequence, any initial difference between two density matrices $\rho^1_0$ and $\rho^2_0$ vanishes over successive steps.  For example, one may track the trace distance between two states
\begin{equation}
D(\rho^1_k, \rho^2_k) = \mathrm{Tr} \bigl|\rho^1_k - \rho^2_k\bigr|
\end{equation}
as a function of the step index $k$.

In Ref.~\cite{Sannia2024dissipationas}, it was shown that for reservoir computing based on Markovian quantum dynamics, ESP can be ensured by choosing the time‐evolution step appropriately.  This result rests on the fact that trace distance monotonically decreases under any Markovian quantum channel.  Conversely, a non‐Markovian quantum channel may violate the ESP.

\subsection{Numerical check of ESP}
Regarding the reservoir state updated by Eq.~(\ref{eq:qrc-res}), we tested asymptotic convergence to the same state by repeatedly applying the update starting from two different initial density matrices:
\begin{align}
\rho_\mathrm{max} &= \frac{1}{2^N} I,\\
\rho_\mathrm{sep} &= \ket{00\cdots0}\bra{00\cdots0}.
\end{align}
We set $N=7$, $N_\mathrm{sys}=4$, and $N_\mathrm{env}=3$, with all other parameter settings as in the STM task.  As reservoir observables, we used the single‐qubit expectation values $Z_i$.

In Fig.~\ref{fig:esp_7_4_3}, panels (a) and (b) show the evolution of the squared norm of the difference between the reservoir‐state vectors for two different random seeds—i.e., two different realizations of the Hamiltonian interaction coefficients. Both Figure (a) and Figure (b) illustrate the state differences in the Markovian region (blue), the intermediate region (green), and the non-Markovian (red) region. Please note that the vertical ($y$-axis) scale varies between the figures.

For both seeds, under strong non-Markovianity, the squared norm of the difference does not fully decay.  In other words, when starting from two distinct initial states, the reservoir states fail to asymptotically converge to the same point even after long evolution, suggesting an effective breakdown of the ESP.  Under strong non-Markovianity, depending on the specific realization of the interaction coefficients, the difference may exhibit large oscillations around small values (as in (a)) or slowly oscillating and decaying around a large value (as in (b)).

Looking at (a) and (b), under strong Markovianity the squared norm remains arbitrarily close to zero, whereas in the “Intermediate regime,” moderate non-Markovianity gives rise to identifiable non-zero regions.  The seed-dependent variations observed here are expected to be alleviated by scaling up the size of the quantum system.

From the above, it is suggested that quantum reservoirs with strong non-Markovianity may severely violate the ESP, implying that conventional linear‐regression‐based readouts may not achieve stable training and inference; consequently, more complex nonlinear readouts should be considered \cite{kubota2024reservoircomputinggeneralized, Ohkubo2024GenSyncRC, attention_ro}.  Moreover, due to the non-stationarity associated with non-Markovianity, the conventional ESP may not apply, and an ESP formulation for non-stationary reservoir updates should be investigated \cite{Kobayashi_ESP}.

\section{Performance in Time Series Prediction Benchmark}
\subsection{NARMA Series Prediction Task}
NARMA (Nonlinear Auto-Regressive Moving Average) series are a representative time-series prediction benchmark in which the next value is generated by nonlinearly combining long histories of past outputs and inputs. In this study, we employ the NARMA-$n$ sequence, which allows for stepwise evaluation of memory length and degree of nonlinearity. In general, NARMA-$n$ is defined by the following recurrence:
\begin{equation}
    y_k = a y_{k-1} + b y_{k-1} \frac{1}{n} \sum_{i=1}^n y_{k-i} \ + c u_{k-n}u_{k-1} + d . 
\end{equation}
Here, following common convention, we set $(a, b, c, d) = (0.3, 0.05, 1.5, 0.1)$. Also, to prevent overflow during computation, the second term is divided by $n$. $y_k$ is the output at time $k$, and $u_k\in[0,0.5]$ is the input sequence generated by a uniform random number. To inject $u_k$ to the quantum reservoir, we applied min-max scaling to $u_k$ to obtain $s_k$. Since the calculation of the output $y_k$ requires the past $n$ steps of output history $\{y_{k-1},\dots,y_{k-n}\}$ and input history $\{u_{k-1}, \dots, u_{k-n}\}$, the model must maintain long-term dependencies. Furthermore, because the output and inputs are coupled by products and higher-order input terms (e.g., $u_{k-n}u_{k-1}$ or $u_k^3$), a high degree of nonlinear transformation capability is required. In this study, we evaluate performance on held-out validation data using the R$^2$ score between the true output $\{y^\mathrm{(val)}_k\}_{k>L_\mathrm{train}}$ and the corresponding model output $\{\hat{y}^\mathrm{(val)}_k\}_{k>L_\mathrm{train}}$. We consider NARMA tasks of orders $n=1,5,10,20,30,40,50$.

\begin{figure*}[t]
\centering
\includegraphics[scale=0.60]{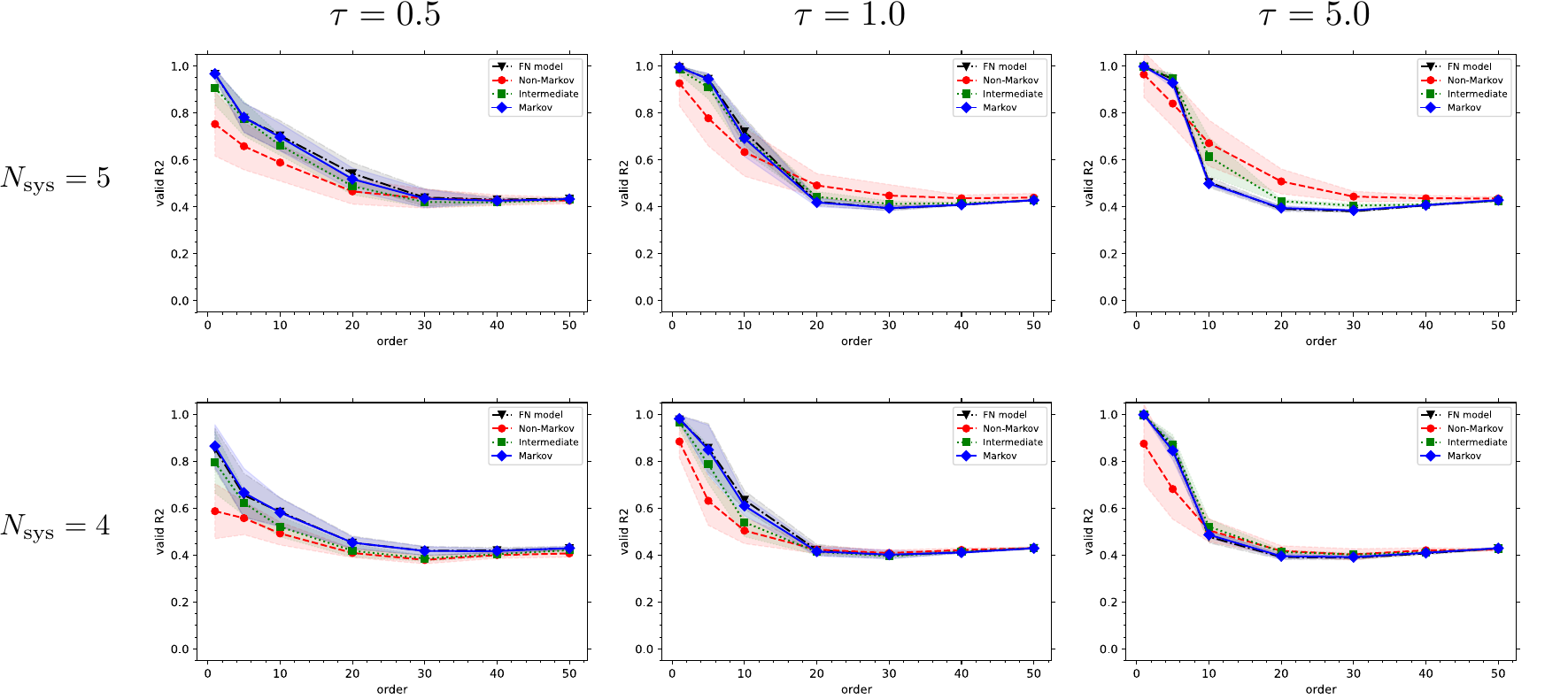}
\caption{$R^2$-score on validation data for the NARMA task. The black line represents the FN model with $N_{\rm env}=0$, and the other lines show the proposed model with $N_{\rm env}\neq0$, depicting the evolution of the $R^2$-score. The upper panel is for $N_{\rm sys}=5$, and the lower panel for $N_{\rm sys}=4$. For the proposed model, $N_{\rm env}$ is set to 2 in the upper panel and 3 in the lower panel, so that $N=7$ in both cases.}
\label{fig:high_narma}
\end{figure*}

\subsection{Results}
Figure~\ref{fig:high_narma} presents the validation-data $R^2$-scores for the NARMA tasks of orders $n=1,5,10,20,30,40,50$.  The parameter sets $(\alpha, \beta) = (5.0, 0.1), (0.1, 5.0), (1.0, 1.0)$ were used to adjust the strengths of the intra-environment and system–environment interactions, corresponding respectively to the “Markov regime” (blue line), “Non-Markov regime” (red line), and “Intermediate regime” (green line).  The black line represents the case $N_\mathrm{env}=0$, hereafter referred to as the FN model (Fujii–Nakajima model) \cite{Fuji-Naka}.  In the upper panel (excluding the FN model), we set $N_\mathrm{sys}=5$ and $N_\mathrm{env}=2$, while in the lower panel we set $N_\mathrm{sys}=4$ and $N_\mathrm{env}=3$.  The intra-system coupling strength was fixed at $J_0=1$, and the magnetic fields were set to $h^\mathrm{(sys)}=J_0$ and $h^\mathrm{(env)}=\alpha J_0$.  The time-evolution step width was chosen as $\tau=0.5, 1.0, 5.0$ (left, middle, right), and the number of virtual nodes for time multiplexing was $V=20$ \cite{Fuji-Naka}.  As reservoir observables, we used both the single-qubit expectation values $Z_i$ and the two-qubit correlations $Z_i Z_j$.  The lines and shaded regions denote the mean and standard deviation, respectively, over ten different random seeds.

Notably, in all panels of Figure~\ref{fig:high_narma}, the $R^2$-score curves for the “Markov regime” and the FN model almost coincide.  This suggests that by adding environment qubits and incorporating non-Markovian dissipation, one can achieve memory characteristics and nonlinear transformation capabilities that differ from those of the original model.  Furthermore, for $\tau=5.0$ with $N_\mathrm{sys}=5$, the decay of the $R^2$-score as a function of the NARMA order is more gradual, and in particular for NARMA-20 and NARMA-30 the “Non-Markov regime” exhibits the highest explanatory power.


\section{Conclusion}
In this work, we have introduced a first-principles, Hamiltonian-driven framework for engineering non-Markovian quantum reservoirs. By explicitly specifying the system, environment, and interaction Hamiltonians—and tuning their coupling strengths and spectral properties—we demonstrated a seamless interpolation between Markovian and non-Markovian dynamical regimes. Our numerical studies on short-term memory benchmarks revealed that reservoirs with stronger non-Markovianity retain information over longer timescales, exhibiting a fragmented form of long-term memory and a markedly slower decay in the $R^2$-score. Through an analysis of the echo-state property, we showed that strong environment coupling can violate the usual contraction condition, giving rise to persistent oscillations in the reservoir state trajectories. Finally, our exploration of higher-order NARMA tasks confirmed that, by adjusting the time-evolution step size, non-Markovian reservoirs can outperform their Markovian counterparts on complex temporal processing benchmarks.

These results highlight the power of Hamiltonian engineering to harness environmental memory effects in analog quantum hardware platforms—such as coupled spin networks, superconducting circuits, and cold-atom lattices—in a resource-efficient manner. Looking ahead, we plan to investigate scalable implementations in near-term devices, optimize reservoir architectures via spectral shaping, and explore hybrid classical–quantum learning pipelines. 

We anticipate that our approach will open new avenues for building versatile, high-performance quantum reservoir computers capable of tackling a wide range of time-series and signal-processing tasks.

\vspace{10pt}
\section{Additional Information}
During the preparation of this manuscript, a related study was published by another research group, which also explores the exploitation of non-Markovian effects in quantum reservoirs. In their groundbreaking work \cite{sannia2025nonmarkovianitymemoryenhancementquantum}, the authors introduce tunable non-Markovianity into otherwise Markovian quantum reservoirs by interleaving partial-SWAP entangling gates with depolarizing channels acting on ancillary qubits. While this approach represents a significant advance in the controlled injection of memory effects, our work instead focuses on a purely Hamiltonian-based framework tailored to analog quantum-physical reservoir computing. By specifying and engineering the system, environment, and interaction Hamiltonians directly, we provide a complementary and resource-efficient route to harness non-Markovian dynamics in platforms such as coupled spin networks, superconducting circuits, and cold-atom lattices.
\section{Acknowledgments}
This work was partially supported by the  New Energy and Industrial
Technology Development Organization (NEDO) under grant number JPNP23003 and the Japan Science and Technology Agency (JST), Advanced Technologies for Carbon-Neutral (ALCA-Next) Grant Number JPMJAN24B1

\bibliographystyle{unsrt}  
\bibliography{references}

\end{document}